%% file: paper.tex
\documentclass[10pt, conference, compsocconf]{IEEEtran}
\usepackage{url}
\usepackage{listings}
\usepackage{fancyvrb}
\usepackage{framed}
\usepackage{graphicx}
\usepackage{todonotes}

\lstnewenvironment{mycode}[1][]
  {\lstset{#1,
           frame=single, 
           captionpos=b,
           basicstyle=\footnotesize\ttfamily}}
  {}

\begin{document}

%%%
%%% title
%%%
\title{Deriving program transformations by demonstration}

%%%
%%% authors
%%%
\author{\IEEEauthorblockN{Matthew J. Sottile}
\IEEEauthorblockA{Galois, Inc.\\
Portland, OR, USA\\
Email: mjsottile@gmail.com}
\and
\IEEEauthorblockN{Geoffrey C. Hulette}
\IEEEauthorblockA{Sandia National Laboratories\\
Livermore, CA, USA\\
Email: ghulette@gmail.com}
}

\maketitle

%%%
%%% input sections
%%%
\input{abstract}

\input{intro}

\input{related}
\input{methods}
\input{algorithms}

\input{conclusion}

%%%
%%% bib
%%%
\bibliographystyle{plain}
\bibliography{paper}

\end{document}

%% file: abstract.tex
\begin{abstract}

Automatic code transformation in which transformations are tuned for specific
applications and contexts are difficult to achieve in an accessible manner. In
this paper, we present an approach to build application specific code
transformations. Our approach is based on analysis of the abstract syntax
representation of exemplars of the essential change to the code before and
after the transformation is applied.  This analysis entails a sequence of
steps to identify the change, determine how to generalize it, and map it to
term rewriting rules for the Stratego term rewriting system. The methods
described in this paper assume programs are represented in a language- neutral
term format, allowing tools based on our methods to be applied to programs
written in the major languages used by computational scientists utilizing high
performance computing systems.

\end{abstract}

%% file: intro.tex
\section{Introduction}

Automated program transformation is a method for changing programs for the
purposes of porting, improving, and maintaining code.  Such techniques are
attractive because they can help prevent the introduction of bugs due to
mistakes and the inefficiency of manual, repetitive changes to potentially huge
source code bases. In the context of high performance computing, the necessity
to perform extensive changes to programs has been brought to the front of 
developers minds due to the rapid change in architecture and corresponding
programming models in petascale and emerging exascale systems. Any assistive
technology to aid developers using those systems would improve their working
lives.

The practice of refactoring~\cite{fowler99refactoring}, wherein behavior-
preserving changes are applied to code, has been well established. This is due
in large part to the inclusion of refactoring algorithms in popular tools such
as integrated development environments (IDEs). These tools may be inadequate,
however, when developers wish to customize transformations for 
application-specific purposes. 
Tools such as Stratego~\cite{visser01stratego} that take a term rewriting
approach to program transformation encourage customization, providing a domain
specific rewriting language for expressing and composing transformations.  The
cost of this is the exposure of complex rewriting systems that are beyond the
scope of knowledge (and often patience) of many programmers. Refactoring tools
within IDEs hide this low level pattern matching and term manipulation from
the user.  Our research focuses on term
rewriting-based tools, aiming to reduce the cost of entry to creating
transformations.  Our goal is to allow rewrite system rules to be generated in
a semi-automated fashion with guidance from the user, insulating them as much
as possible from the underlying term representation and rewriting mechanics.

The work described in this paper has been inspired by a popular tool called
Coccinelle\footnote{\url{http://coccinelle.lip6.fr/}}. Coccinelle is used for
describing and performing application-specific transformations on C code. The
technique that we describe expands upon the concepts introduced by Coccinelle in
three ways.

\begin{enumerate}

\item{\bf Broad language support:} we seek to support all languages that are
commonly used in computational science in addition to C, such as Fortran and
C++.

\item{\bf Native language transformation specification:} we use the original
source language to specify the transformation using compiler directives and
code annotations to guide the transformation generation.  Coccinelle  uses a
C-like domain specific language called SmPL for transformation specification.

\item{\bf Structural difference driven rule generation:} we employ a different
algorithm for inferring the rewriting rules that are derived from the
transformation specification based on previous work in the area of structural
tree comparison for difference identification.  The rules that we generate
target the Stratego term rewriting system.
\end{enumerate}

\subsection{Motivating example}

Consider the following transformation as a motivating example. Changing the way
in which data is laid out is a relatively simple optimization that can result in
significant performance benefits on modern massively multithreaded
architectures~\cite{stratton12optimization}. The performance benefits are
realized through improved data locality in the transformed code. A common
transformation pattern for this sort of optimization transposes a structure of
array-based fields into an array of structures containing singleton elements
for each field. Implementing this kind of transformation is conceptually simple,
but in practice can be quite tedious. In particular, implementations will
require modification of at least the following points within the program: data
structure definition, allocation, deallocation, and both direct and indirect
access to fields. Even though transformations of the structure definition
are easily accomplished by hand, accesses to instances of the
structure will appear throughout the source code and will require repetitive
application of one or more transformations.  These repetitive transformations are
the motivator for automation.

\subsection{Approach}

\begin{figure}
\centering
\includegraphics[width=0.3\textwidth]{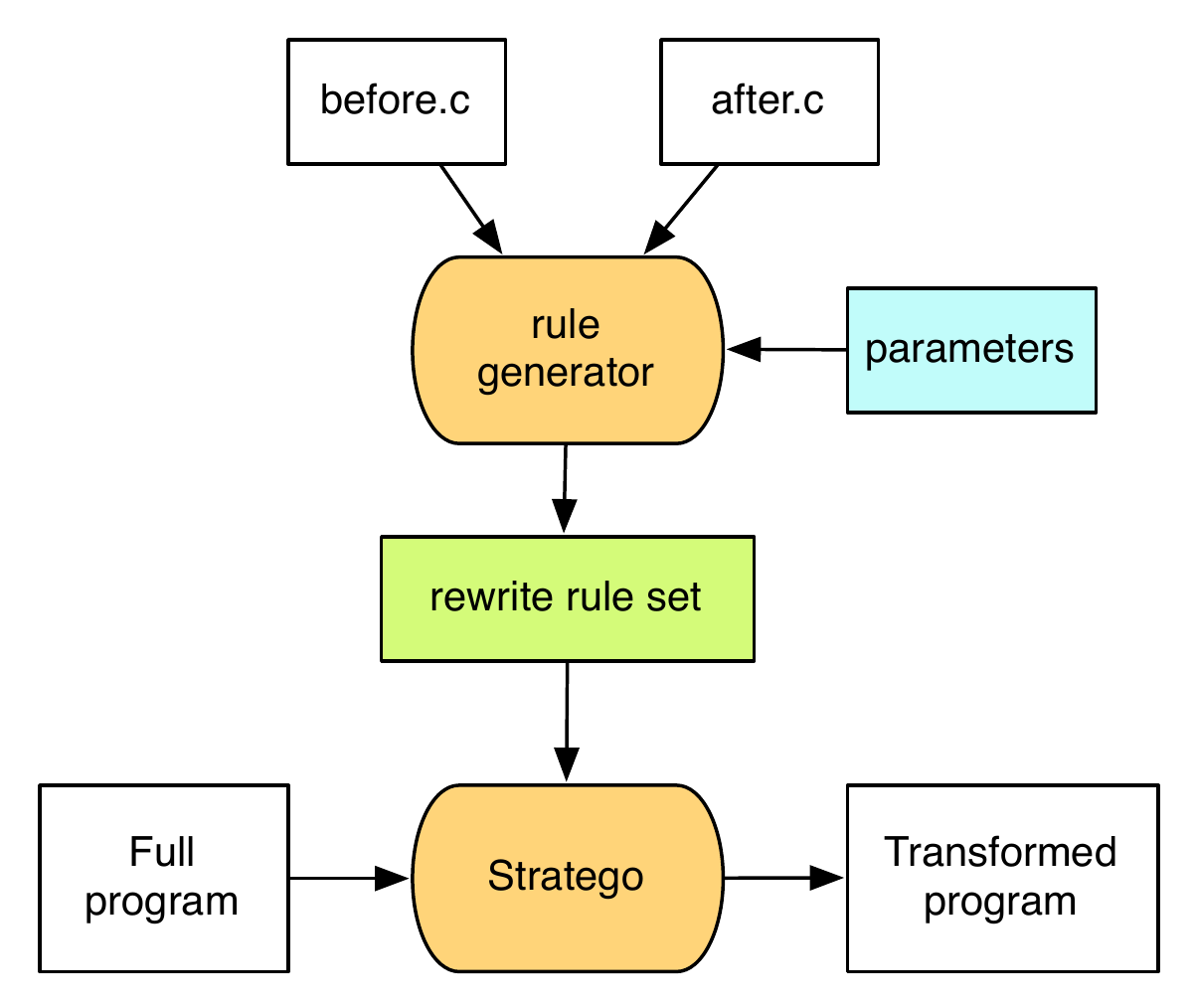}
\caption{Demonstration-based workflow.}
\label{fig:workflow}
\end{figure}

Our approach uses a demonstration-based development workflow illustrated
in Figure~\ref{fig:workflow}. Programmers
write small programs that represent the ``before'' and ``after'' states of
essential parts of the code under the desired transformation. Our tool infers
from these demonstrations a set of rewrite rules that will carry out the
transformation. The inference operates on a term representation derived from
the abstract syntax tree (AST) of the code. Our technique uses a structural
differencing algorithm to determine the set of tree edit operations necessary
to carry the term representation of the code before the transformation to the
term representation of the code after. The rewrite rules inferred by this
process are parameterizable by heuristic-driven refinements. Thus, our
approach allows for rules to be generalized into patch-like changes, and also
to include context around changes that can control where the rule is applied.
In this paper we will describe our methodology and algorithmic approach, as
well as known limitations of our current implementation and areas of current
and ongoing research and  development.

%% file: related.tex
\section{Related work}

In the last decade, a number of customizable code transformation tools arose
based on formal term rewriting methods~\cite{baader99term}. Within the context
of program transformation, term rewriting is applied by mapping an abstract
representation of program code to a general term representation that can be
manipulated by a term rewrite system. A generic rewriting engine that
implements a specific rewriting system is then applied to combine the term
representation of the program with a set of rewriting rules that capture the
steps to implement in the overall program transformation.  These techniques,
while quite powerful and general, impose a significant burden on the
programmer to learn the details of the rewriting system --- expression of
transformations as rewrite rules in a system like Stratego~\cite{stratego} or
Maude~\cite{maude} is challenging. This paper addresses this problem of
defining these transformations in a way that reduces the necessary familiarity
with the underlying rewrite technology and term representation used to perform
the actual transformation.

The concept of program transformation based on some high-level specification
is not new. Our work is directly inspired by the
Coccinelle~\cite{andersen08generic} program transformation tool for C code.
The Coccinelle system uses the concept of a \emph{semantic patch} to represent
transformations using a format similar to that used by the familiar UNIX {\tt
patch} tool. The semantic patch specification language
(SmPL~\cite{padioleau06smpl}) that Coccinelle uses provides a combination of
familiar C language syntax combined with additional information to identify
metavariables, share metavariables between rules, and sequence rule
application. Similar information is required by our tool to guide steps such
as transformation generalization and context definition, and we are working to
integrate it via structured comments in the language used to specify the
before and after transformation code that are input to our tool.  In the
interim, our prototype implementation requires heuristic parameters to be
specified separate from the before and after code specifications.  The reason
for this is that it allows us to use existing parsing and analysis
infrastructure (such as ROSE) to support relevant languages for scientific
programmers, with extraction of annotations performed as an analysis phase on
the parse tree or AST.

The HERCULES~\cite{kartsaklis12hercules} project states very similar goals as
the work described in this paper.  HERCULES focuses on specifying code
patterns and transformations in such a way that they are accessible to the
programmer.  HERCULES also exposes compiler optimizations to the user of the
tool, allowing them to include information to be used at compile time in the
tuning process.  Our work focuses on the specification of transformations via
a patch-like definition of the code before and after the transformation has
taken place. We rely on information provided via parameters (and eventually
code annotations) similar to those that HERCULES specifies via compiler
pragmas to communicate additional information to the transformation tools that
cannot be inferred from the patch specification alone.  Information about how
the programmer expects the patch to be generalized, or to provide context to
limit its scope of application, are examples of our use of annotations.  The
methods that we describe in this paper would complement systems such as
HERCULES in adding automation to the process of pattern and transformation
specification to lower the cost of entry for programmers unfamiliar with the
inner workings of complex compilers or pattern matching systems.

%% file: methods.tex
\section{Methodology}

Our approach is based on the user providing a minimal input representing a
specific example of a change that they wish build a rewrite rule for. Changes
are expressed by specifying before and after code snippets that are an example
of the change to capture.  Adopting a format modeled on the UNIX {\tt patch}
tool, we can compactly view the before and after snippets where lines prefaced
with a minus sign ({\tt -}) exist only in the before code, lines prefaced with a
plus sign ({\tt +}) exist only in the after code, and all other lines appear in
both. For example, if one wants to derive a rule to apply the distributive law
for multiplication over addition, this could be expressed as:

\begin{mycode}[caption=Distributive law of arithmetic.]
void foo() {
  int x,y,z,a;

-  x = a*(y+z);
+  x = a*y + a*z;
}
\end{mycode}

The goal of the before/after versions of the code is to provide a minimal
example that embodies only the change that a rewrite rule should be generated
for.  We are investigating methods to use annotations (either in the form of
compiler pragmas or structured comments) that can communicate additional
information that can be used by the rule generation algorithm.  In the interim
this information exists as parameters provided to our prototype tool.  This
additional information is necessary to aid in rule generalization, defining
appropriate context for changes to be defined within, and defining
relationships between multiple changes that result in a sequence of rules that
must be applied in a specific sequence. 

From this demonstration of the change(s), our algorithm infers one or more
rewrite rules that implement the change in a generalized form.  For example,
a rewrite rule generated by our prototype of the algorithms described in
this paper for the distributive law example is:

\begin{mycode}[caption=Distributive law rewrite rule inferred from examples.]
R1 : multiply_op(
        T_1, add_op(T_2, T_3, T_4, _),
        T_4, _) 
     -> 
     add_op(
        multiply_op(T_1, T_2, T_4, gen_info()),
        multiply_op(T_1, T_3, T_4, gen_info()),
        T_4, gen_info())
\end{mycode}

A rewrite rule is defined by two patterns (the left hand side to match and the
right hand side to replace it with) separated by an arrow ({\tt ->}).  In this
case, the rule {\tt R1} has on its left hand side a term that represents a
tree rooted at a multiplication operation, with two children - an arbitrary
expression {\tt T\_1} and a tree rooted at an addition operator. The addition
operator itself has two children {\tt T\_2} and {\tt T\_3} which represent
arbitrary expressions.  Additional structure (such as the {\tt T\_4} element)
is a consequence of the AST representation that carries additional information
related to typing and source locations.  On the right hand side of the rule,
we see the addition operator has been promoted to the root of the tree, with
the multiplication distributed to the children appropriately.  The bare
underscores that appear in the pattern on the left hand side of the rule
correspond to arbitrary subterms that are disregarded and replaced with the
special {\tt gen\_info()} term.  This term is used by the tool that maps the
terms back to an AST representation to generate source locations for the AST
elements created or moved as part of the transformation.

\subsection{Algorithmic stages}

Given an example in which the before and after states of the transformation
are expressed, the algorithm for computing rewrite rules is structured
in a sequence of steps.

\begin{enumerate}

\item{\bf Term Generation:} Code is mapped to a term representation using the
Annotated Term (aterm) format~\cite{brand00aterm} used by Stratego/XT. We use
the Minitermite tool included with the ROSE compiler framework for term
generation.

\item{\bf Structural difference calculation:} Identification of structural
differences between the two examples in their term representation.  This yields
a patch that can be applied to implement \emph{precisely} the change that was
present in the examples.

\item{\bf Difference generalization:} Examples are written in terms of
specific program elements (e.g., variable and function names). Generation of
more generally applicable structural patterns entails the introduction of
metavariables to be used during pattern matching by the rewrite engine.
Metavariables act as named ``spaces'' in the terms in which arbitrary legal
subterms can reside, and can be referred to by the metavariable name in
rewrite rule patterns.

\item{\bf Context introduction:}. Establishment of context allows the rewrite
engine to distinguish common term substructures in order to apply the rule at
the appropriate place in the tree. For example, function argument lists appear
in both declarations and function calls: a transformation may be intended to
only apply at call sites will require added context to the argument list to
include parent nodes in the AST that represent the function call site.  This
eliminates unintended pattern matches and rule applications at function
declarations.

\item{\bf Rewrite rule generation:} Creation of Stratego rewrite rules. This
phase combines traversals of the structures produced by the structural
difference computation with information from annotations and parameters that are
necessary to relate rules and control their order of application.

\end{enumerate}

The second through fourth steps are the core of this work.  The second step is
focused on determining precisely what structures are present in both the before
and after transformation representation of the program.  The abstract
representation of a program often contains more explicit detail about the
structure of a program than the plaintext representation that the programmer
works directly with.  By using the source representation to express the code
at both ends of a transformation step, the programmer is insulated from the
abstract representation, and the relevant structures can be extracted
automatically.

The third step is important in deriving abstract patterns to match from the
concrete examples provided by the programmer defining the before and after
transformation structure. For example, consider the code snippets discussed
earlier expressing the distributive law.  Looking only at the structural
difference between the terms representing these examples, we find that the
variables in the expression are bound to specific variable instances ($a$,
$x$, $y$, and $z$). The rewrite rules that implement this change derived
directly from the source code would apply only for this exact expression - any
other example with other variable names, sub-expressions, or arithmetic
operators would fail to pattern match in the rewrite engine.

On the other hand, the programmer may intend to generate a transformation that
represents application of the distributive law for expressions of this form
with arbitrary legal structures in place of concrete variables (e.g., function
calls, sub-expressions, etc...).  In this case, we would like to perform a
refinement step on the terms corresponding to the structural difference
between the examples to replace concrete named variables with metavariables
that represent arbitrary legal terms within the arithmetic expression.
Generalization is essentially the act of replacing specific AST node instances
with named holes in which any legal AST structure can occur.

The fourth step addresses a similar problem, but involving the parent and
higher nodes in the tree relative to the detected change.  Generalization is
primarily concerned with children of the nodes where changes appeared.  The
parents of a changed node are necessary to establish context such that the
scope of matching for the rule is constrained to the subtrees where the
change is meaningful.  For example, introduction of a statement requires us to
define a pattern for the portion of the AST where the code does not yet exist,
but will after the rule is applied.   This pattern will include the parent
node as well as some set of sibling nodes nearby the change in order to give
the rewrite system a frame of reference for finding precisely where we wish to
make the change.  As we will discuss in more detail in
Section~\ref{sec:context}, this choice is not well defined and the manner by
which we define what constitutes sufficient context to define a rewrite rule
pattern is a component of our ongoing research work.

\subsection{Term generation}

Everything that we describe in this paper is based on our ability to map
programs to and from a term representation that can be analyzed to infer
rewrite rules, as well as fed into a rewriting engine such as Stratego in
order to apply these rules to transform code.  We achieve this term mapping
via the ROSE compiler framework\footnote{\url{http://www.rosecompiler.org/}}
and a tool called Minitermite originally developed as part of the SATIrE
(Static Analysis Tool Integration Engine)
project~\cite{DBLP:conf/isola/BaranyP10}. Minitermite is able to traverse the
ROSE Sage AST and map AST nodes to and from a term representation. By taking
this approach, we are able to leverage the fact that ROSE supports numerous
languages in its Sage AST format and handles the challenging task of parsing
and generating code in each supported language. By using ROSE and Minitermite,
we are able to bypass much of the tedium of the language front-end and code
generation process and focus our algorithms on transformations applied to the
generic Sage AST in term form.

%% file: algorithms.tex
\section{Algorithms}

In this section we will discuss the algorithms that implement the stages of
the rule generator.  During processing, the representation of the code changes
depending on the operations being performed on it.  These representations
correspond to data types used to represent the tree structure, which are
defined in Haskell for this paper and our corresponding implementation.  The
initial term representation that is used for the before and after versions of
the program is simply a labeled tree in which the root of the term is stored
as a label, and subterms are stored as a list of children.

\begin{verbatim}
data LabeledTree = 
    Node String [LabeledTree]
\end{verbatim}

The algorithm for computing the edit distance between the two trees
requires more information to indicate not only what resides within the
tree but what edit operations occur at the nodes.

\begin{verbatim}
data EditTree = 
    ENode String [(EditOp,EditTree)]
  | ELeaf LabeledTree

data EditOp = Keep | Delete
\end{verbatim}

The result of the computation is a pair of edit trees representing the sequence
of operations necessary to turn each tree into the other.  In order to generate
rewrite rules that represent the changes between the sides, we must establish a
relation between the elements of each edit tree.  This relation between each
node within the edit tree represents one of four interpretations of a pair of
edit operations on a node: the node matches between the two trees, the node does
not match, the node is not present in the ``before'' tree but is in the
``after'' tree, and the node is not present in the ``after'' tree but is in the
``before'' tree. Adopting the convention that ``before'' and ``after''
correspond to the left- and right-hand sides of the rule respectively, we
represent the absence on one side or the other as a hole indicating the side.

\begin{verbatim}
data WeaveTree = 
    WNode Label [WeavePoint]
  | WLeaf LabeledTree

data WeavePoint = 
    Match WeaveTree
  | Mismatch WeaveTree WeaveTree
  | LeftHole WeaveTree
  | RightHole WeaveTree
\end{verbatim}

It is important to note that edit operations are computed from the root
towards the children.  When a mismatch or hole is detected, no further
comparison is attempted below that point.  This is why both edit and weave
tree nodes have a leaf case from which we hang the original labeled subtree.

\input{structdiff}
\input{generalize}
\input{contextualize}
\input{rulegen}

%% file: structdiff.tex
\subsection{Structural difference calculation}

Rewrite systems operate on graph structures, most often in the form of trees
representing the abstract syntax of programs.  As such, naive string
differencing algorithms (such as edit distance) are not easily applied to
compare structured data.  We instead started our work using algorithms for
computing an edit distance between two tree structures derived from the
abstract syntax representation of the code.  From the tree distance
computation we obtain both an edit distance metric as well as a sequence of
edit operations that can be performed to the given trees to transform one into
the other.  The choice of algorithms for computing such a difference is broad,
as the need for reasoning about changes in structured data arises in many
contexts beyond program AST understanding such as comparing XML
documents~\cite{tekli09overview}.

In this work, our goal is to determine for two trees the simplest set of
operations necessary to map one to the other.  We restrict ourselves to the
simple set of edit operations: add, delete or keep.  The algorithm that we
implemented represents additions implicitly as a hole on one side and a delete
operation that is paired with the hole on the other side.  Richer sets of edit
operations have been studied for representing tree edit distances (such as
whole-subtree movement), which often can be represented as a sequence of the
simpler add/delete/keep operations.  The basis of our work is the algorithm
presented by Yang~\cite{yang91diff} that was used to visualize source
differences where the differences were informed by the syntactic structure of
the programs.  

Yang's algorithm is similar to previously described algorithms
by Tai~\cite{tai79tree} and Selkow~\cite{selkow77tree}, with allowances for
the concept of ``comparable symbols''.  The ability to support comparable
symbols allows difference calculations to be made more or less sensitive to
parts of the abstract syntax tree that can be considered interchangeable.  The
use of these flexible comparison operators is an aspect of our ongoing
research work.  For example, a transformation on a structure that requires
precedence of operators to be respected regardless of which operators are present
could benefit from an comparison operator that treats binary operators of equal
precedence as equivalent.

% Yang's algorithm is a straightforward dynamic programming algorithm.  The code
% for our implementation is shown in Listing~\ref{fig:yang-code}.

\subsection{Weaving edit trees}

Once we have obtained two {\tt EditTree} structures from the algorithm, we
then wish to determine how the two trees related such edit operations can be
associated with the paired before/after AST objects.  This algorithm differs from
that presented by Yang, as Yang's work was only concerned with printing the
difference between the programs and not maintaining the computed edit
structure for further computation.  Our algorithm takes the two {\tt EditTree}
structures and yields a single {\tt WeaveTree}, named such due to its role as
representing the two edit trees as essentially overlain and woven together to
form a unified tree of edit operations.  
%In Figure~\ref{fig:editToWeave}, we
%show a very simple example of a pair of edit trees for a pre/post pair, and
%the resulting weave tree.

\begin{table*}[tb]
\centering
\begin{tabular}{|r|l|l|l|}
\hline
   & {\bf Pre-EditTree}     & {\bf Post-EditTree}   & {\bf Woven Tree}                      \\ \hline
1. & {\tt []}               & {\tt []}              & {\tt []}                              \\
2. & {\tt (Keep tL):restL}    & {\tt (Keep tR):restR}   & {\tt (Match (weave tL tR)):weaveHandle restL restR} \\
3. & {\tt (Delete tL):restL}  & {\tt (Delete tR):restR} & {\tt (MisMatch tL' tR'):weaveHandle restL restR}    \\
4. & {\tt (Delete,t):rest}    & {\tt []}              & {\tt (RightHole t'):weaveHandle rest []}          \\
5. & {\tt []}               & {\tt (Delete,t):rest}   & {\tt (LeftHole t'):weaveHandle [] rest}           \\
6. & {\tt (Delete tL):restL}  & {\tt r}               & {\tt (RightHole tL'):weaveHandle restL r}         \\
7. & {\tt l}                & {\tt (Delete tR):restR} & {\tt (LeftHole tR'):weaveHandle l restR}          \\
8. & {\tt []}               & {\tt (Keep t):rest}     & {\bf error}                           \\
9. & {\tt (Keep t):rest}      & {\tt []}              & {\bf error}                           \\ \hline
\end{tabular}
\caption{Weaving operations for edit tree node pairs.  The function {\tt weaveHandle} recursively implements this table.  Subtrees with a prime ({\tt '}) annotation correspond to unmodified LabeledTree instances that are attached to the Woven Tree.}
\label{table:weaveops}
\end{table*}

Weaving a pair of nodes together requires consideration of the table of
possible pairings as shown in Table~\ref{table:weaveops}.  The cases are
rather straightforward to break down.  The base case (1) states that weaving
two empty lists is itself empty.  Case 2 handles the situation where we have
two matching nodes indicated by paired \emph{Keep} edit operations.  This
results in a \emph{Match} node being created in which the children of the
matching nodes are woven together, followed by the remaining siblings of the
matched node.  Case 3 is the opposite of this in which both nodes are not the
same have a \emph{Delete} edit operation indicating a mismatch.  A
\emph{MisMatch} node is created in that case, with the two deleted subtrees
attached, followed by the siblings of the mismatching nodes being processed.

Case 4 represents the deletion of an element from the before code when no
nodes remain in the after code.  This occurs when deleting elements from the
end of a list, where the list on the before side would be longer.  This
results in a \emph{RightHole} node being created indicating that the right-
hand side of the comparison was missing nodes.  Case 5 is the symmetric
instance of this where the deletion operation appears in the after code,
resulting in a \emph{LeftHole} being created.  Cases 6 and 7 are similar,
except they appear when the side without the deletion operation still have
elements remaining.  An example of this occuring would be the deletion or
insertion of an element in a list at a postion before the end. Finally, cases
8 and 9 are the symmetric error cases that are impossible to encounter in
practice but are included to complete the set of patterns to match.  Both
cases represent the situation in which one side is the empty list, yet the
other side has a \emph{Keep} operation.  \emph{Keep} operations must be paired
on both sides since they represent matches between the two trees being
compared.  Clearly it is not possible match a concrete element with an empty
set, therefore we assert an error if these cases are hit.

%% file: generalize.tex
\subsection{Term generalization}

\begin{figure}
\centering
\includegraphics[width=0.4\textwidth]{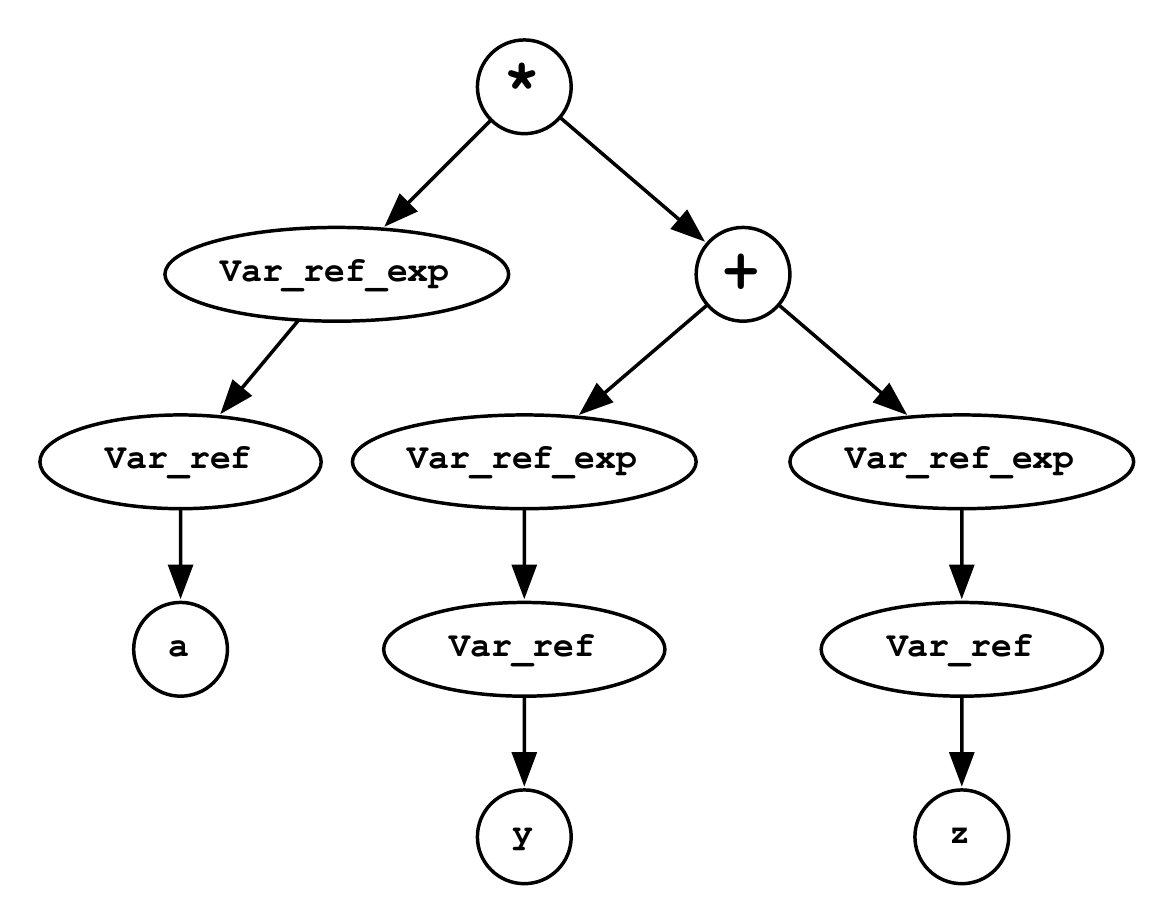}
\caption{Term without variable reference generalization.}
\label{fig:pre-no-generalize}
\end{figure}

The result of computing the structural difference between two programs is a
sequence of edit operations in which subtrees are deleted, inserted, or
replaced. These edit operations correspond specifically to the ASTs of the
input programs and do not generalize without additional processing.  Consider
again the simple case of the distributive law in which the pre-transformation
example is the expression {\tt x = a * (y + z)}.  When the right hand side of
the assignment is identified as the location of the change by the structural
differencing algorithm, we would see a term that has a structure similar to
the tree shown in Figure~\ref{fig:pre-no-generalize}.  This tree will match
not only the desired arithmetic expression structure, but will also require a
match to contain the exact variable references as well.

\begin{figure}
\centering
\includegraphics[width=0.2\textwidth]{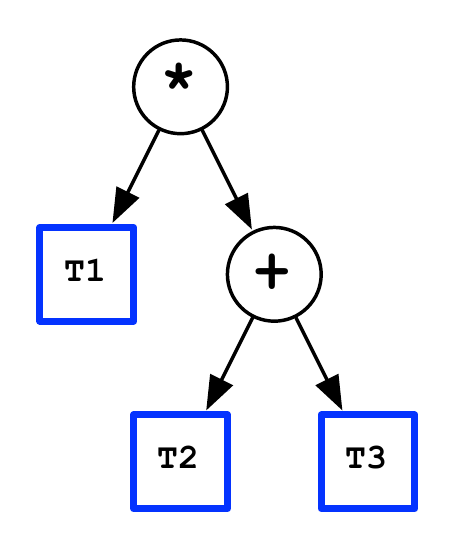}
\caption{Term with variable reference generalization via metavariable introduction.}
\label{fig:pre-with-generalize}
\end{figure}

On the other hand, if we use heuristics that define term
processing rules such as ``all variable reference expressions should be
replaced with meta-variables'', then we can create a tree that represents a
more general pattern as shown in Figure~\ref{fig:pre-with-generalize}.  This
generalized pattern then allow any legal subterm to match in place of the metavariabes.

Generalization of patterns for rule generation presents two tasks to solve: the algorithm to
process the terms and replace substructures with generic patterns, as well as the
specification rules that represent legitimate generalizations.  

\subsubsection{Generalization algorithm}

The algorithm that we have implemented approaches generalization by
establishing terms to seek out in which specific subterms are replaced with
metavariables.  For example, consider the case where we are seeking to
generalize arithmetic expressions to build a pattern that matches on the
operator structure and is oblivious to the specific operands (e.g., variable
references or function calls).  This could be implemented by matching all
subtrees that are rooted at one or more arithmetic operators ({\tt
multiply\_op}, {\tt add\_op}, and so on).  For each subtree that matches these
roots, we could then seek out further subtrees contained within them that are
rooted at terms that we wish to replace, such as {\tt var\_ref\_exp} and {\tt
binary\_op\_annotation} nodes. These nodes correspond to specific variables
being referenced, or metadata that ROSE uses internally to indicate
information such as the inferred type of a binary operator.  In both cases, if
we replace these subtrees with metavariables, then when the original binary
operator that triggered the search for those terms is matched, the pattern
matcher will focus on the operator structure of the expression and allow
arbitrary legal operands to be matched.

\subsubsection{Term processing specification}

We currently specify a generalization as a pair $g = (R, S)$ where $R$ is a
set of labels corresponding to the roots of subtrees that we wish to
traverse in the interest of generalizing.  The set $S$ is the set of
labels corresponding to terms within subtrees rooted at an element of $R$ 
that we wish to replace with metavariables.  All elements of $R$ are
treated as equivalent in their interpretation during generalization.
Generalizations are specified via parameters independent of the before/after
code, and are represented within our prototype as an ordered sequence
$G = (g_1, g_2, \cdots)$ that dictates their order of application.
This allows certain generalizations to be applied before others in the event
that the order of application matters.  Enforcing the order ensures that their
application will be predictable.

%% file: contextualize.tex
\subsection{Context generation}
\label{sec:context}

In addition to meta-variable introduction, we also require the introduction of
context in the rules.  This is most apparent when considering transformations
that do not replace AST with new AST parts, but introduce new AST from nothing
or delete AST without replacement. Common examples of this include the
addition or deletion of a parameter to a function call, statement within a
block, or else-clause within a conditional.  Context can also be used to
control the scope of application for rules that could be applied in
undesirable places absent context.

Consider the following case in which a status argument is to be added to a
function call.  This involves the introduction of the status variable
declaration as well as the inclusion of an additional parameter in the
function invocation.

\begin{mycode}[caption=Addition of a variable declaration and function parameter.]
void bar(void) {
   int x,y,z;
+  int a;

-  x = foo(y,z);
+  x = foo(y,z,a);
}
\end{mycode}

\begin{figure*}[tb]
\centering
\includegraphics[width=0.8\textwidth]{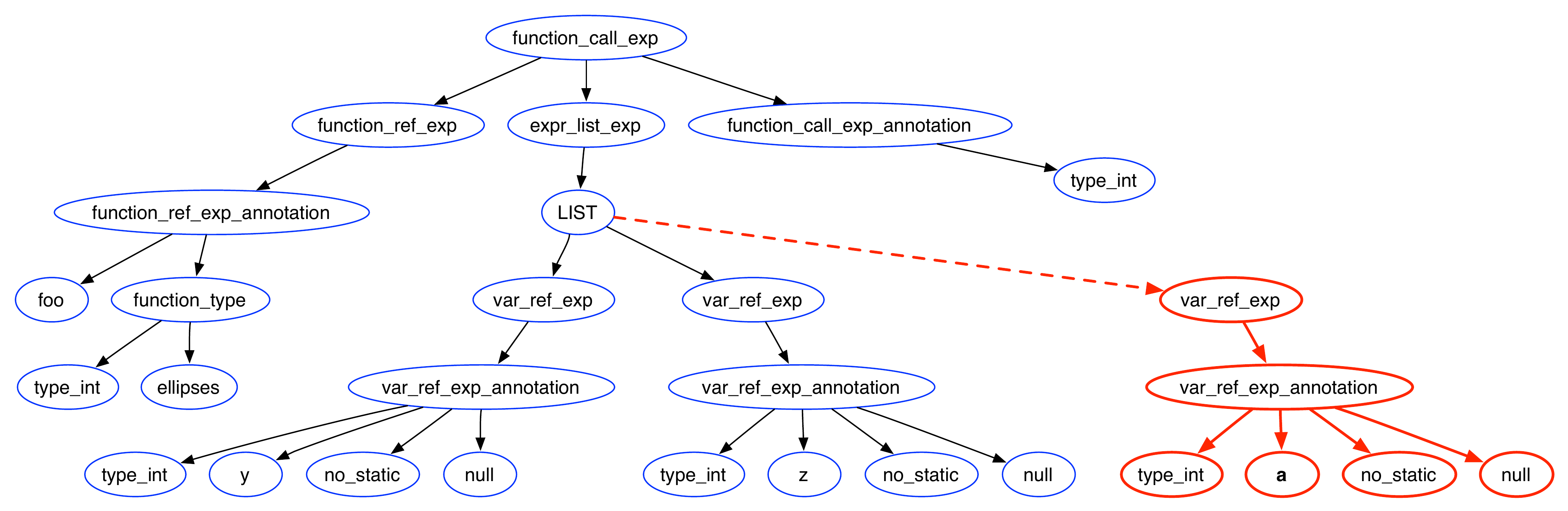}
\caption{Addition of an argument to a function call.  The added AST nodes are indicated with
the dashed line and red nodes.}
\label{fig:context-add-argument}
\end{figure*}

The differencing algorithm will correctly identify that an insertion was
performed by identifying a hole in the input tree.  Unfortunately, to build a
usable rewrite rule, the hole must be given context to allow a pattern to be
defined that can be matched.  This context can be found by looking at the
parents of the AST where the hole appears.  The challenge is that the number
of AST nodes towards the root of the tree that are required for sufficient
context requires some thought.  Consider the tree shown in 
Figure~\ref{fig:context-add-argument} in which a single argument 
called {\tt a} is added to
the parameter list for the function call to {\tt foo()}.

\begin{figure}
\centering
\includegraphics[width=0.4\textwidth]{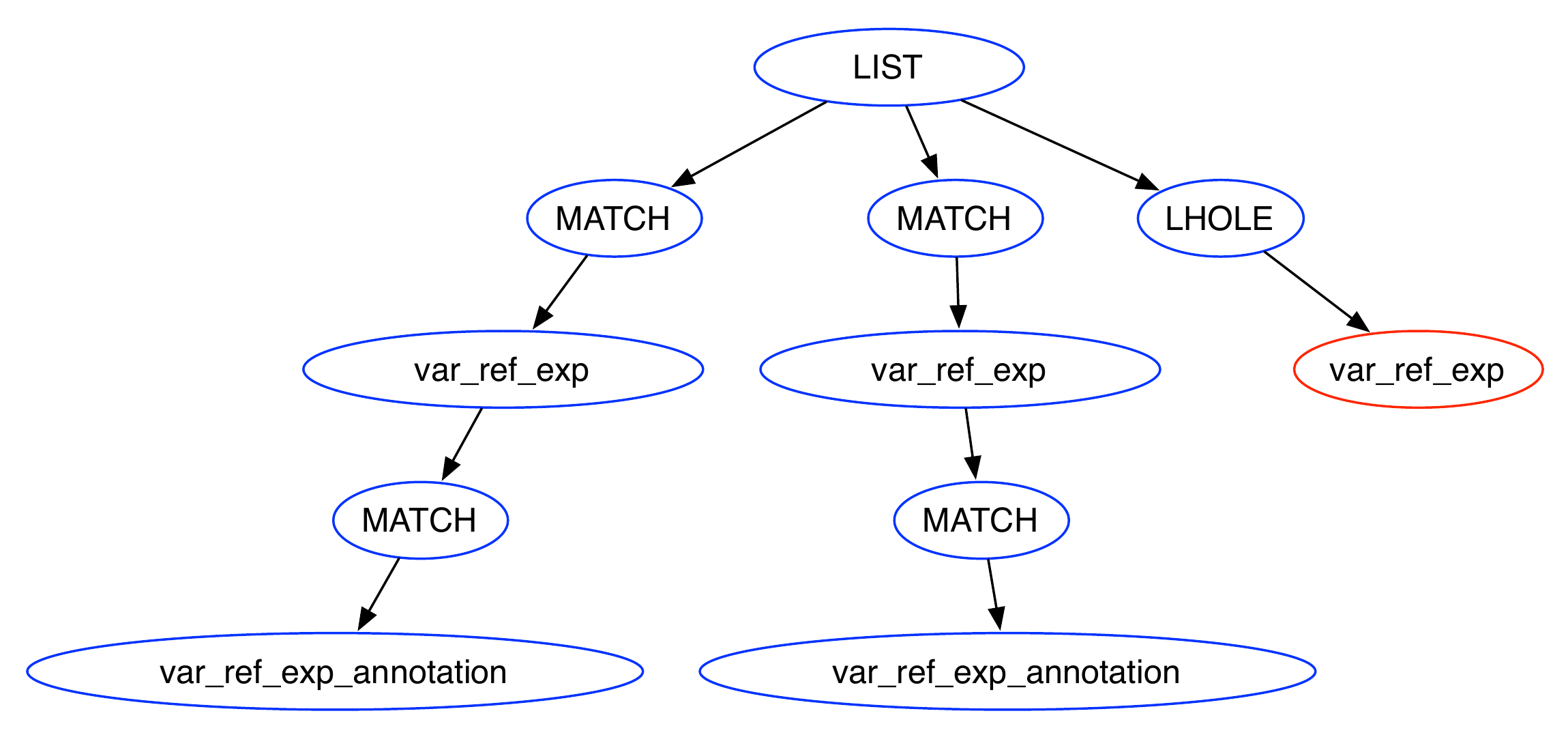}
\caption{Illustration of the hole appearing in the woven edit trees.  This list corresponds
to the {\tt expr\_list\_exp} in Figure~\ref{fig:context-add-argument}.}
\label{fig:context-weave}
\end{figure}

In this case, we see that the variable reference expression added to the list
of arguments is indicated by the dashed line.  This line originates from the
aterm list node that represents the list of expressions that form the argument
list for the function call expression at the root.  When traversing the woven
tree, the insertion appears as a left-hole indicating that something present
in the right hand tree (the post tree) is paired with an absent element in the
left hand tree.  This is illustrated in Figure~\ref{fig:context-weave}, in
which we see the match weave points for the two arguments that are common to
both the pre and post versions of the code.

The question that we are faced with is determining how far up the tree is
necessary to build context in order to control where this pattern is matched.
For example, without context the rule would simply state that in the absence
of any term, insert the variable reference expression.  This is clearly wrong,
as it is ambiguous and could lead to a proliferation of insertions all over
the AST.  If we look up at the parent of the inserted AST elements, we see the
aterm list.  This is slightly better, but still will cause uncontrolled
insertion of variable reference expressions all over the program wherever a
list containing the other two parameters appears.  As we move up the tree, we
include more context that narrows down the set of potential pattern matches
that will be found by the rewrite engine.  In this case, if we are interested
in only rewriting function calls, we need to traverse from the hole until we
hit an ancestor that is a function call expression.

Specification of context is currently a work in progress.  Our current
prototype traverses the weave tree from the root and for subterms whose root
is in a prespecified set of labels of interest (e.g., function declarations),
we perform further processing.  This processing involves determining whether
or not these subterms contain hole nodes, indicating that code is removed or
deleted and requires context to be added to the rule.  While this has proven to
be useful to generate legitimate rule patterns with sufficient context, we
ultimately wish to drive this process not from a set of subtree roots to seek, but
information provided with the before/after code specification by the user.  This
would be consistent with the method used by Coccinelle in the SmPL language.

%% file: rulegen.tex
\subsection{Stratego rewrite rule generation}

The bulk of the work related to rule generation is performed in the
previous steps, such as the replacement of terms with metavariables and 
pairing of pre/post terms that correspond to the left and right hand
side of rewrite rules.  The set of term constructors that are necessary
for establishing the structure of terms that Stratego will work with
will be provided by Minitermite.  These term constructors define the
legal structure of terms, which must match the structure that Minitermite
produces from the ROSE Sage AST.  The Sage AST contains a great number 
of nodes, with a tiny representative subset shown below for illustrative
purposes.

\begin{verbatim}
signature
  sorts E F A
  constructors
    gen_info    : F
    file_info   : S * N * N -> F
    add_op      : E * E * A * F -> E
    multiply_op : E * E * A * F -> E
\end{verbatim}

These term constructors establish the set of term sorts that terms are
composed of.  For example, both {\tt gen\_info} and {\tt file\_info}
correspond to term constructors that map AST constructs to concrete
source locations, both of which yield the sort {\tt F}.  The binary
operators {\tt add\_op} and {\tt multiply\_op} represent AST nodes
that appear in arithmetic expressions (with sort {\tt E}).  We can see
that the binary operators have associated with them terms of sort {\tt F}
since they have some location within the source files.  The full set of
sorts and constructors is independent of any specific rule set, but serves
as a common definition that are used for any transformation rules generated
by our algorithms.

Given this boilerplate, term generation currently is implemented as a simple
traversal of the AST terms identified (with generalization and context)
by the structural differencing algorithm.  The traversal serializes the
structures as strings that are aggregated in a Stratego {\tt .str}
specification file.  This file is then compiled with the Stratego compiler
to yield an executable that consumes programs to be transformed in
aterm form, yielding another aterm that Minitermite can then reconstruct
as ROSE Sage AST nodes for code generation.

Future work in Stratego rule generation includes potential use of
Stratego strategies in order to relate rules and control their order of
application.  We are specifically looking at how to use program annotations
and other directives provided by the user to generate this additional
information in the Stratego rule sets beyond that which can be derived from
the processed structural differencing computation.

%% file: conclusion.tex
\section{Conclusion}

Automatic code transformation techniques that can be driven by application
programmers with minimal knowledge of the underlying transformation and
code analysis tools are very important for maintaining and evolving
complex simulation codes.  The results and techniques that we show in this
paper build upon techniques that have been developed for the C language
community, but focus on applying them to languages that are commonly
used in high performance and scientific computing.  Our prototype implementations
of these techniques have been able to be applied to simple cases to demonstrate
their utility in lowering the cost of entry for programmers to use automated
transformation tools.  We have a number of ongoing lines of research related to
this work to address questions that arise in supporting more complex and nuanced
transformations than discussed in this paper.

\subsection{Current work and prototype}

This paper discusses the core algorithms used in our prototype.  The major areas
of ongoing work at submission time include:

\begin{itemize}

\item Moving parameters currently specified via configurations separate
  from the before/after code specification into code annotations either via
  structured comments or compiler directives.

\item Techniques for specifying points in the ancestors of holes to identify
  the required context for transformations.  This will augment or fully
  replace our current heuristic approach.

\item Use of Stratego strategies to coordinate the application of multiple rules
  that constitute a single complex transformation.

\end{itemize}

All of the tools and techniques used in this work are available as open source
software available via the source control repository at
\url{http://sf.net/projects/compose-hpc/}.  We hope that interested readers
will try our evolving prototypes, and contributions to advance the work are
always welcome.

\subsection{Acknowledgements}

This work was supported by the U.S. Department of Energy Office of Science, contract no.
DE-SC0004968.  We would like to thank Adrian Prantl for his assistance with the ROSE-based
Minitermite source-to-term and term-to-source conversion tools that were critical in
mapping between source code in a variety of languages and the generic aterm representation
that our algorithms operate on.